\documentclass[journal]{IEEEtran}
\usepackage{amsmath,amssymb,bm}
\usepackage{booktabs,multirow,tabularx,array}
\usepackage{algorithm}
\usepackage{algpseudocode}
\usepackage{tikz}
\usetikzlibrary{positioning,arrows.meta,shapes.geometric,fit,calc,backgrounds}
\usepackage{pgfplots}
\usepackage{pgfplotstable}
\usepgfplotslibrary{groupplots}
\pgfplotsset{compat=1.18}
\usepackage[caption=false,font=footnotesize]{subfig}
\usepackage{cite}
\usepackage{url}
\usepackage[hidelinks]{hyperref}
\usepackage{xcolor}
\usepackage{graphicx}
\usepackage{enumitem}
\usepackage{microtype}
\setlist[itemize]{leftmargin=*,nosep}
\newcommand{\TrustAUPRC}{0.342}
\newcommand{\TrustBrier}{0.088}
\newcommand{\TrustECE}{0.082}
\newcommand{\TrustCorr}{0.222}
\newcommand{\TrustAuto}{0.317}
\newcommand{\TrustMIA}{0.499}
\newcommand{\AUPRCGain}{7.5\%}
\newcommand{\BrierReduction}{6.1\%}
\newcommand{\ECEReduction}{17.0\%}
\newcommand{\CorrReduction}{28.0\%}
\newcommand{\AutoReduction}{30.1\%}

\newcolumntype{Y}{>{\raggedright\arraybackslash}X}
\begin{document}
\title{Differentially Private and Fairness-Audited Score Diffusion for Irregular Longitudinal Health Records}
\author{Taimoor Ahmed \thanks{Superior University Lahore, Pakistan}}\maketitle
\begin{abstract}
Sharing irregular longitudinal health records can accelerate model development, yet synthetic releases may leak participation, distort temporal dependence, suppress rare events, or reduce utility for underrepresented groups. We present TRUST-LONGSYNTH, an auditable patient-level private generator that combines bounded sufficient statistics, zCDP-accounted Gaussian releases, conditional analytic score diffusion, block-banded temporal covariance, separate missingness and gap models, and a protected-event sampling floor with population weights. The method was evaluated on five independently generated, three-cohort benchmarks containing 720 patients, fourteen irregular observation slots, six mixed variables, informative missingness, and a rare deterioration outcome. At $\varepsilon=12$ and $\delta=10^{-5}$, TRUST-LONGSYNTH achieved mean train-synthetic-test-real AUPRC 0.342, Brier score 0.088, expected calibration error 0.082, correlation error 0.222, autocorrelation error 0.317, and membership-attack AUROC 0.499. Relative to the private diagonal score baseline, AUPRC increased by 7.5\%, while Brier, calibration, correlation, and autocorrelation errors decreased by 6.1\%, 17.0\%, 28.0\%, and 30.1\%, respectively. The method did not dominate every nonprivate or discrete baseline, and corrected paired tests were inconclusive with five seeds. Canary exposure was 1.8\%, compared with 28.8\% for DP-Score in the same stress test. These findings support a transparent privacy-utility-fairness evaluation protocol, not clinical validity or unconditional release safety, and motivate governed external validation on real multi-site records.
\end{abstract}
\section{Introduction}
Longitudinal electronic health records and remote-monitoring streams contain clinically useful temporal structure, but they are difficult to share because a patient is represented by a sequence of correlated observations rather than one independent row. A generator that reproduces realistic marginals while copying rare trajectories, suppressing uncommon outcomes, or changing the relationship between measurements and time can create both privacy and scientific risks. Public critical-care resources such as MIMIC-IV and eICU have enabled reproducible research under explicit governance arrangements~\cite{johnson2023,pollard2018,a4,a6}, while systems such as Synthea generate fully simulated records from a population model~\cite{walonoski2018,a9,a10}. Neither route removes the need for methods that can transform governed, irregular, institution-specific trajectories into data products whose privacy boundary, utility, and failure modes are inspectable.

Synthetic health-data research has progressed from discrete visit generation and tabular adversarial models to temporal generative networks, variational methods, and diffusion models~\cite{choi2017,yoon2019,lin2020,xu2019,baowaly2019,desai2021,kotelnikov2023}. However, a synthetic dataset is not anonymous merely because it contains fabricated identifiers. Membership inference can exploit differences between training and non-training records~\cite{shokri2017,a7}, canary tests can expose unintended memorization~\cite{carlini2019,a11}, and broad evaluations have shown that high apparent resemblance does not establish safe release~\cite{stadler2022,a12}. Clinical generators have demonstrated useful downstream performance and, in some cases, formal private training~\cite{beaulieu2019,tucker2020,a92}, yet longitudinal reviews still report fragmented handling of temporal dependence, irregularity, privacy, and evaluation~\cite{perkonoja2026,miletic2026}. Accordingly, privacy, statistical fidelity, task utility, and subgroup behavior must be assessed jointly rather than treated as interchangeable scores.

Differential privacy (DP) provides a mathematical relation between outputs produced from neighboring datasets~\cite{dwork2006,dworkroth2014}. Its value is strongest when the unit of adjacency matches the protected entity. For longitudinal health data, that unit is normally a patient, not an observation. Modern accountants based on concentrated or R\'{e}nyi DP can compose multiple Gaussian releases more tightly than naive addition of approximate-DP parameters~\cite{bun2016,mironov2017}; neural approaches commonly use clipped, noisy gradients~\cite{abadi2016}. Nevertheless, formal accounting alone does not establish usefulness. Large noise can erase rare outcomes or temporal associations, and a favorable average can conceal poor utility for an underrepresented group. These concerns are especially important in health applications, where algorithmic disparities may reinforce existing inequities~\cite{rajkomar2018,obermeyer2019}.

The technical problem is therefore multi-objective. Let one patient contain mixed measurements, missingness masks, unequal time gaps, a cohort or site label, an approved group variable used only for auditing, and a rare downstream outcome. The generator must bound each patient's contribution; release all data-dependent parameters through an auditable privacy mechanism; preserve enough conditional and temporal structure to train useful predictors; avoid collapsing a small protected-event stratum; and expose empirical attacks and fidelity diagnostics without presenting any one diagnostic as a proof of safety. Existing private tabular methods such as PrivBayes and PATE-GAN address important portions of this problem~\cite{zhang2014,jordon2019}, but static rows do not directly represent irregular sequences. Conversely, temporal GANs and diffusion models provide expressive sequence or density modeling~\cite{yoon2019,ho2020,song2021} without automatically supplying patient-level privacy or subgroup guarantees.

We address this gap with TRUST-LONGSYNTH, an analytic conditional score-diffusion framework whose complete training-data-dependent parameter bundle is released through patient-bounded Gaussian mechanisms. The model uses broad, prespecified support bounds; clips each patient's encoded trajectory; releases conditional first and second moments; projects the noisy covariance to a positive-semidefinite block-banded matrix; and transports Gaussian noise through an analytic score path. Observation masks and irregular gaps are modeled separately rather than hidden by regular-grid imputation. A protected-event sampling floor ensures that a small stratum is represented in the synthetic training resource, while population-correction weights retain an estimate of the released population distribution. Because sampling is post-processing of the private bundle, repeated generation does not consume additional privacy budget under the stated mechanism.

The study is deliberately organized as a reproducible methodological proof of concept. Five independently generated populations each contain 720 simulated patients, three shifted cohorts, fourteen irregular observation slots, six mixed variables, informative missingness, one underrepresented group, and a rare deterioration outcome. Patient-disjoint training, validation, and test partitions prevent duplicated individuals from crossing evaluation boundaries. Seven generators are compared using train-on-synthetic/test-on-real (TSTR) discrimination and calibration, marginal and temporal fidelity, subgroup utility, membership and attribute attacks, canary exposure, missingness robustness, privacy-budget sensitivity, ablation, and execution cost. General and task-specific utility are reported separately, following the distinction emphasized in statistical-disclosure work~\cite{snoke2018,goncalves2020}.

At the primary privacy setting, TRUST-LONGSYNTH obtained mean TSTR AUPRC \TrustAUPRC, Brier score \TrustBrier, expected calibration error \TrustECE, correlation error \TrustCorr, autocorrelation error \TrustAuto, and membership-attack AUROC \TrustMIA. Relative to the private diagonal score baseline, the corresponding AUPRC gain was \AUPRCGain, while Brier, calibration, correlation, and autocorrelation errors decreased by \BrierReduction, \ECEReduction, \CorrReduction, and \AutoReduction. These differences were directionally consistent for several metrics but did not survive Holm correction with five seeds. The proposed method also did not dominate every comparator: a private Markov baseline and an independent conditional generator achieved higher mean AUPRC. We therefore frame the contribution as an auditable privacy--utility--fairness design and evaluation workflow, not as proof of universal superiority, clinical validity, or unconditional release safety.

The contributions are:
\begin{itemize}
\item A patient-level private conditional score-diffusion mechanism that releases bounded conditional moments, block-banded temporal covariance, observation-mask statistics, and gap statistics under one explicit zCDP ledger.
\item A protected-event synthesis rule with population-correction weights and a joint evaluation protocol covering TSTR utility, temporal fidelity, calibration, subgroup outcomes, privacy attacks, canary exposure, and robustness.
\item A complete executable research package containing five-seed results, sensitivity and ablation studies, Python inspection plots, TikZ/PGFPlots paper figures, generated tables, tests, validation scripts, and a governed real-cohort transition schema.
\end{itemize}

The remainder of this paper is organized as follows. Section~II synthesizes related generative, private, and evaluation literature. Section~III defines the longitudinal record, privacy boundary, and optimization objectives. Section~IV presents the private releases, analytic score sampler, auxiliary models, and algorithms. Section~V describes the benchmark, baselines, metrics, results, limitations, and reproducibility checks. Section~VI concludes the paper and identifies the validation required before any clinical or release-safety claim.

\section{Related Work}
\subsection{Clinical and Temporal Synthetic Data}
Early clinical generators represented a patient as a set or vector of codes. medGAN combined an autoencoder with an adversarial generator for multi-label visit records~\cite{choi2017}, while later clinical GAN variants improved optimization and downstream evaluation~\cite{baowaly2019}. These systems established that synthetic records should be tested through clinical prediction tasks, not only visual similarity. However, visit-vector formulations compress timing and often treat missing observations as preprocessing details.

Temporal generators explicitly model ordered sequences. TimeGAN combines adversarial and supervised losses in a learned latent space~\cite{yoon2019}; DoppelGANger jointly generates metadata and regular time series~\cite{lin2020}; and TimeVAE provides a lower-complexity latent-variable alternative~\cite{desai2021}. Diffusion models now offer stable likelihood-free generation through iterative denoising~\cite{ho2020,song2021}, and TabDDPM adapts this principle to mixed tabular columns~\cite{kotelnikov2023}. These advances motivate the score-based component of our method. Nevertheless, regular arrays and independent rows do not directly express a patient's unequal observation gaps, feature-specific masks, and rare outcome label under one patient-level privacy unit.

\subsection{Formal Privacy and Empirical Disclosure}
Private data generation spans Bayesian networks, private teacher ensembles, and noisy neural optimization. PrivBayes releases a private dependency model over discretized attributes~\cite{zhang2014}; PATE-GAN applies noisy aggregation across teacher discriminators~\cite{jordon2019}; and private clinical generators demonstrate that formal privacy can coexist with useful aggregate analyses~\cite{beaulieu2019}. DP-SGD extends privacy to flexible neural models but makes clipping, accountant selection, convergence, and hyperparameter access part of the privacy analysis~\cite{abadi2016}. TRUST-LONGSYNTH instead uses a compact set of patient-bounded sufficient statistics so that every source-dependent release can be listed in a human-readable ledger.

Formal DP and empirical attacks answer different questions. DP constrains output distributions for neighboring inputs, whereas membership attacks search for exploitable behavior in a particular implementation and threat model~\cite{shokri2017}. Canary exposure probes whether an unusual planted pattern is reproduced~\cite{carlini2019}. Stadler \emph{et al.} show why resemblance, nearest-neighbor distance, or a weak attack cannot alone justify an anonymity claim~\cite{stadler2022}. Our evaluation therefore reports the accountant, membership attack, attribute inference, and canary test together, but does not interpret an attack AUROC near 0.5 as proof of safe release.

\subsection{Utility, Fairness, and Evaluation}
Utility is multidimensional. General-purpose measures assess marginals, correlations, transitions, and temporal dependence; task-specific measures test whether synthetic training supports a predefined downstream objective~\cite{snoke2018,goncalves2020}. In healthcare, subgroup evaluation is also necessary because average performance can hide unequal errors or diminished utility for a population already disadvantaged by data coverage~\cite{rajkomar2018,obermeyer2019}. Recent reviews confirm that longitudinal synthesis studies use inconsistent datasets, temporal metrics, privacy diagnostics, and validation designs~\cite{perkonoja2026,miletic2026}. Accordingly, our protocol fixes the split unit, downstream event, group audit, privacy budget, fidelity metrics, and attack settings before comparing methods.

\begin{table*}[t]
\caption{Closest method families. DP: formal differential privacy; Irr.: explicit irregular timing/missingness; SG: subgroup audit; Att.: empirical attacks.}
\label{tab:related}
\centering\scriptsize
\setlength{\tabcolsep}{3.1pt}
\begin{tabular}{p{2.55cm}p{2.0cm}ccccp{5.0cm}}\toprule
Study & Core model & Temp. & DP & Irr. & SG/Att. & Limitation relative to this work \\ \midrule
medGAN~\cite{choi2017} & Autoencoder+GAN & -- & -- & -- & Att. & Visit-vector synthesis; no patient-level private irregular sequence model. \\
TimeGAN~\cite{yoon2019} & Supervised temporal GAN & Yes & -- & No & -- & Preserves dynamics but has no formal patient-level privacy accounting. \\
DoppelGANger~\cite{lin2020} & Metadata+sequence GAN & Yes & -- & Partial & -- & Strong regular time-series model; privacy remains empirical. \\
CTGAN~\cite{xu2019} & Conditional tabular GAN & No & -- & No & -- & Mixed columns, but rows are not irregular patient trajectories. \\
medWGAN/\allowbreak medBGAN~\cite{baowaly2019} & Clinical GAN variants & Partial & -- & No & -- & Clinical utility tests without a formal release boundary. \\
Private clinical GAN~\cite{beaulieu2019} & DP neural generator & Partial & Yes & No & Limited & Demonstrates private trial synthesis; no explicit gap/mask or worst-group design. \\
PATE-GAN~\cite{jordon2019} & Private teacher ensemble & No & Yes & No & -- & General tabular private synthesis, not irregular longitudinal health data. \\
PrivBayes~\cite{zhang2014} & Private Bayesian network & No & Yes & No & -- & Auditable, but static discretized dependencies limit continuous trajectories. \\
TimeVAE~\cite{desai2021} & Temporal VAE & Yes & -- & No & -- & Low-rank temporal generation without formal privacy. \\
TabDDPM~\cite{kotelnikov2023} & Tabular diffusion & No & -- & No & -- & Stable mixed-tabular synthesis, but no patient sequence or privacy guarantee. \\
Stadler et al.~\cite{stadler2022} & Privacy evaluation & General & -- & -- & Att. & Shows empirical anonymity claims can fail; not a generator. \\
Perkonoja et al.~\cite{perkonoja2026} & Systematic review & Yes & Mixed & Mixed & Mixed & Finds fragmented longitudinal evaluation; does not provide one mechanism. \\
\textbf{This work} & Private analytic score diffusion & Yes & Yes & Yes & SG+Att. & Auditable joint mechanism; still requires governed real-cohort validation. \\ \bottomrule
\end{tabular}
\end{table*}

Table~\ref{tab:related} exposes the central gap. Expressive temporal generators generally lack a patient-level formal release boundary, while established private tabular methods do not jointly model mixed irregular trajectories, masks, gaps, rare events, and subgroup utility. Private clinical generators narrow this gap but commonly integrate irregularity through undisclosed imputation or do not publish a release-by-release accountant. Evaluation studies identify disclosure or utility failures but do not provide a mechanism that links those diagnostics to model components.

TRUST-LONGSYNTH therefore focuses on a missing systems property rather than adding another unaccounted architecture: every training-dependent parameter used at sampling time must appear in the privacy ledger. The block-banded covariance targets temporal dependence; separate private auxiliary models target missingness and irregular timing; and protected-event sampling targets small-stratum representation. The experiments are designed to test whether each component changes the metric it is intended to affect. Because related methods and journal scopes continue to evolve, this gap statement remains a candidate novelty claim pending a submission-month systematic search and full-text review.

\section{System Model and Problem Formulation}
\subsection{Entities, Records, and Assumptions}
The protected database contains $N$ patients. Patient $i$ has at most $T$ observation slots and $V$ mixed variables. A slot is not assumed to be equally spaced: $\Delta_{it}$ is elapsed time since the preceding slot, and $M_{itv}$ states whether variable $v$ was observed. The approved audit attribute $G_i$, rare downstream outcome $Y_i$, and cohort/site $C_i$ are modeled jointly because their distribution affects both subgroup utility and membership risk. One patient record is
\begin{equation}
\mathcal{R}_i=\left(\mathbf{X}_i,\mathbf{M}_i,\boldsymbol{\Delta}_i,G_i,Y_i,C_i\right),
\label{eq:record}
\end{equation}
where $\mathbf{X}_i\in\mathbb{R}^{T\times V}$, $\mathbf{M}_i\in\{0,1\}^{T\times V}$, and $\boldsymbol{\Delta}_i\in\mathbb{R}_{+}^{T}$. The fixed-slot representation is a computational container, not an equal-time assumption; the gap vector preserves elapsed time. The method assumes prespecified public support bounds for each variable, a fixed adjacency definition, an approved use of $G_i$ for auditing, and a prespecified outcome $Y_i$. It does not assume that a synthetic record corresponds to a real person.

A compact condition vector encodes the main and selected interaction effects needed for rare-event synthesis:
\begin{equation}
\mathbf{c}_i=[1,G_i,Y_i,\mathbb{1}(C_i=1),\mathbb{1}(C_i=2),G_iY_i,Y_i\mathbb{1}(C_i=2)]^{\top}.
\label{eq:condition}
\end{equation}
The interactions are intentionally limited to avoid unstable private estimation. More complex cohort or intersectional effects require larger governed samples and a revised privacy allocation.

\begin{table*}[t]
\caption{Principal notation.}
\label{tab:notation}
\centering\footnotesize
\begin{tabular}{p{1.7cm}p{5.6cm}p{1.7cm}p{5.6cm}}\toprule
Symbol & Meaning & Symbol & Meaning \\ \midrule
$N,T,V$ & Patients, slots, and variables & $\mathcal{R}_i$ & Complete patient-level record \\
$\mathbf{X}_i$ & Mixed longitudinal values & $\mathbf{M}_i$ & Observation mask \\
$\boldsymbol{\Delta}_i$ & Unequal time gaps & $G_i,Y_i,C_i$ & Audit group, event, and cohort \\
$\mathbf{c}_i$ & Bounded condition vector & $\mathbf{z}_i$ & Encoded and clipped value trajectory \\
$B_v$ & Public support interval for variable $v$ & $L$ & Patient-level $\ell_2$ clipping radius \\
$\rho$ & zCDP privacy parameter & $(\epsilon,\delta)$ & Approximate-DP parameters \\
$\widetilde{\mathbf{A}}$ & Private condition Gram matrix & $\widetilde{\mathbf{B}}$ & Private condition--value cross moment \\
$\widetilde{\mathbf{S}}$ & Private value second moment & $\widehat{\boldsymbol\beta}$ & Private conditional-mean coefficients \\
$\widehat{\boldsymbol\Sigma}$ & Stabilized temporal covariance & $b$ & Temporal block bandwidth \\
$\sigma_k$ & Score-path noise level & $q_s,p_s$ & Synthetic and released population stratum mass \\
$w_i$ & Population-correction weight & $\mathcal{L}_{\rm task}$ & Downstream task loss \\
\bottomrule\end{tabular}
\end{table*}

\subsection{Public Encoding and Patient Contribution}
For each variable, the broad support interval $B_v=[a_v,b_v]$ is chosen before inspecting the private database. Observed values are mapped to $[-1,1]$, missing entries are temporarily set to the public midpoint, and gaps are log-scaled. The encoded patient value trajectory is
\begin{equation}
\begin{aligned}
z_{itv}&=M_{itv}\left[2\frac{\operatorname{clip}(X_{itv},a_v,b_v)-a_v}{b_v-a_v}-1\right],\\
r_{it}&=2\frac{\log(1+\Delta_{it})}{\log(1+\Delta_{\max})}-1.
\end{aligned}
\label{eq:encoding}
\end{equation}
Interpolation is used only to form the value-moment query; masks are released through a separate model, so an imputed entry is never presented as an observation. The flattened value vector is clipped as $\bar{\mathbf z}_i=\mathbf z_i\min(1,L/\|\mathbf z_i\|_2)$. This bound is the basis for the sensitivities in Section~IV.

\subsection{Privacy Boundary}
Datasets $D$ and $D'$ are neighbors when one complete patient record is replaced:
\begin{equation}
D\simeq D' \Longleftrightarrow D\setminus\{\mathcal R_i\}=D'\setminus\{\mathcal R'_i\}.
\label{eq:adjacency}
\end{equation}
Replacement adjacency protects the entire sequence, labels, masks, and gaps as one unit. Observation-level adjacency would give a numerically smaller budget but would not match the stated privacy goal.

A randomized mechanism $\mathcal{M}$ is $(\epsilon,\delta)$-DP if, for every neighboring pair and measurable output set $S$,
\begin{equation}
\Pr[\mathcal{M}(D)\in S]\le e^{\epsilon}\Pr[\mathcal{M}(D')\in S]+\delta.
\label{eq:dp}
\end{equation}
For a vector query with $\ell_2$ sensitivity $\Delta_2$, adding isotropic Gaussian noise with standard deviation $\sigma$ provides a zero-concentrated DP contribution
\begin{equation}
\rho=\frac{\Delta_2^2}{2\sigma^2}.
\label{eq:zcdp}
\end{equation}
The ledger adds the $\rho_j$ values of all released queries. For any selected $\delta>0$, the composed mechanism is converted to approximate DP through
\begin{equation}
\rho_{\mathrm{tot}}=\sum_j\rho_j,\qquad
\epsilon=\rho_{\mathrm{tot}}+2\sqrt{\rho_{\mathrm{tot}}\log(1/\delta)}.
\label{eq:composition}
\end{equation}
The implementation allocates this budget before model fitting and rejects any release that would exceed it. Hyperparameter exploration on private validation data would consume additional privacy and is therefore outside the supplied benchmark.

\subsection{Design Objective}
The system seeks a parameter bundle that supports downstream utility while limiting temporal distortion, subgroup degradation, and disclosure evidence. These goals are evaluated separately rather than collapsed into an unverifiable scalar. The formal guarantee applies only to the released bundle and its post-processing; attack metrics characterize the implemented threat models but do not strengthen or weaken Eq.~\eqref{eq:dp}. Similarly, a protected-event sampling rule can improve representation without guaranteeing equal clinical performance. The problem is therefore to construct an auditable mechanism whose intended effects can be tested and whose assumptions remain visible.

\section{Proposed Method}
\subsection{Overview}
\begin{figure*}[t]
\centering
\resizebox{\textwidth}{!}{\includegraphics[]{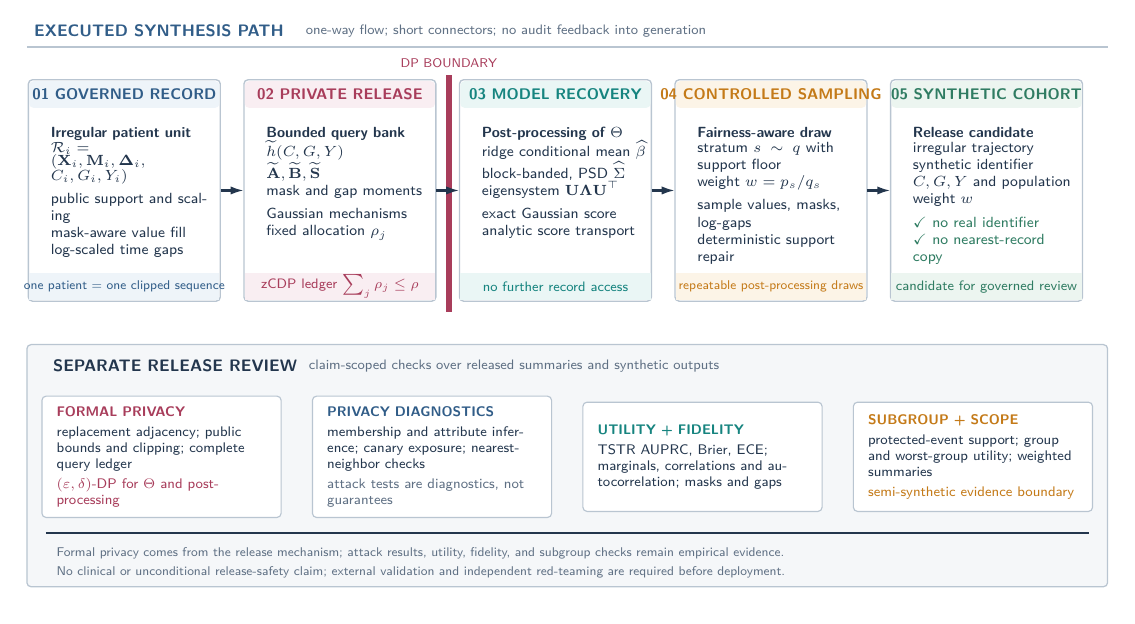}}
\caption{TRUST-LONGSYNTH training and generation. Every database-dependent arrow into the private bundle corresponds to a ledger entry; all operations to the right of the bundle are post-processing.}
\label{fig:architecture}
\end{figure*}

Fig.~\ref{fig:architecture} summarizes the pipeline. Training converts each patient into bounded condition, value, missingness, and gap contributions. A fixed privacy allocation releases the joint condition histogram, condition Gram matrix, condition--value cross moment, value second moment, missingness statistics, and gap moments. Numerical post-processing solves the conditional model, bands and projects the covariance, and creates an analytic score sampler. Generation draws a synthetic condition, transports noise to the corresponding conditional distribution, samples masks and gaps, and emits a population-correction weight. No real identifier or nearest real record is copied into the output.

\subsection{Private Conditional Moments}
Let $\mathbf C$ contain the clipped condition vectors and let $\mathbf Z$ contain clipped encoded value trajectories. The mechanism releases three sufficient statistics,
\begin{equation}
\widetilde{\mathbf A}=\frac{\mathbf C^{\top}\mathbf C}{N}+\mathbf E_A,\quad
\widetilde{\mathbf B}=\frac{\mathbf C^{\top}\mathbf Z}{N}+\mathbf E_B,\quad
\widetilde{\mathbf S}=\frac{\mathbf Z^{\top}\mathbf Z}{N}+\mathbf E_S,
\label{eq:moments}
\end{equation}
where each $\mathbf E$ is independent Gaussian noise calibrated to its query sensitivity and allocated $\rho$. If $\|\mathbf c_i\|_2\le L_c$ and $\|\bar{\mathbf z}_i\|_2\le L$, the replace-one sensitivities are bounded by $2L_c^2/N$, $2L_cL/N$, and $2L^2/N$, respectively. The released conditional-mean coefficients are
\begin{equation}
\widehat{\boldsymbol\beta}=
\left[\Pi_{\mathrm{PSD}}(\widetilde{\mathbf A})+\lambda\mathbf I\right]^{-1}\widetilde{\mathbf B},
\qquad \widehat{\boldsymbol\mu}(\mathbf c)=\mathbf c^{\top}\widehat{\boldsymbol\beta}.
\label{eq:beta}
\end{equation}
The ridge term prevents noisy, weakly supported condition directions from dominating. The same released Gram matrix is reused for auxiliary regressions, which is post-processing rather than another query.

The raw private residual covariance is estimated by subtracting the modeled conditional second moment:
\begin{equation}
\widetilde{\boldsymbol\Sigma}_{0}=
\widetilde{\mathbf S}-\frac{1}{N}
(\mathbf C\widehat{\boldsymbol\beta})^{\top}(\mathbf C\widehat{\boldsymbol\beta}).
\label{eq:covraw}
\end{equation}
Because Gaussian perturbation can make this matrix indefinite and can amplify remote temporal blocks, we apply deterministic block banding followed by eigenvalue clipping,
\begin{equation}
\widehat{\boldsymbol\Sigma}=\Pi_{[\eta,\kappa]}
\left(\mathcal B_b(\widetilde{\boldsymbol\Sigma}_{0})\right),\qquad
[\mathcal B_b(\mathbf Q)]_{tt'}=\mathbf Q_{tt'}\mathbb 1(|t-t'|\le b).
\label{eq:banding}
\end{equation}
Here $\mathbf Q_{tt'}$ denotes a $V\times V$ block. Banding encodes the assumption that nearby slots carry most covariance; it does not claim long-range independence. The bandwidth is fixed from the semi-synthetic design and must be tuned within a privacy-aware protocol on real data.

\subsection{Analytic Score-Diffusion Sampling}
For condition $\mathbf c$ and noise level $\sigma_k$, the perturbed density is Gaussian with covariance $\widehat{\boldsymbol\Sigma}+\sigma_k^2\mathbf I$. Its exact score is
\begin{equation}
\mathbf s_k(\mathbf z\mid\mathbf c)=
-\left(\widehat{\boldsymbol\Sigma}+\sigma_k^2\mathbf I\right)^{-1}
\left[\mathbf z-\widehat{\boldsymbol\mu}(\mathbf c)\right].
\label{eq:score}
\end{equation}
Rather than numerically training a second neural network on private records, the implementation uses the eigendecomposition $\widehat{\boldsymbol\Sigma}=\mathbf U\operatorname{diag}(\boldsymbol\lambda)\mathbf U^{\top}$. A sample at noise level $k$ is transported to $k-1$ by
\begin{equation}
\mathbf z_{k-1}=\widehat{\boldsymbol\mu}+
\mathbf U\operatorname{diag}\!\left(
\sqrt{\frac{\boldsymbol\lambda+\sigma_{k-1}^{2}}
{\boldsymbol\lambda+\sigma_{k}^{2}}}\right)
\mathbf U^{\top}(\mathbf z_k-\widehat{\boldsymbol\mu}).
\label{eq:transport}
\end{equation}
The telescoping endpoint is distributed as $\mathcal N(\widehat{\boldsymbol\mu},\widehat{\boldsymbol\Sigma})$ up to numerical precision. The multi-step path is retained because it exposes the relationship to score diffusion and permits later insertion of constrained or mixed-type correction operators. Medication intensity and oxygen support are rounded or thresholded only after inverse scaling.

\subsection{Irregularity and Small-Stratum Representation}
Missingness is not inferred from imputed values. Let $U_{itv}=1-M_{itv}$. A clipped patient missingness vector and clipped log-gap vector are released through condition cross moments. After ridge post-processing, generation uses
\begin{equation}
\begin{aligned}
\widehat p(U_{itv}=1\mid\mathbf c)&=
\operatorname{clip}([\mathbf c^{\top}\widehat{\boldsymbol\gamma}]_{tv},p_{\min},p_{\max}),\\
r_{it}&\sim\mathcal N([\mathbf c^{\top}\widehat{\boldsymbol\xi}]_t,\widehat\tau_t^2).
\end{aligned}
\label{eq:auxiliary}
\end{equation}
Condition effects are shrunk toward the directly released global mean to control noise in rare strata. A generated trajectory is required to contain at least two observations per variable; the deterministic repair selects the largest mask scores and therefore consumes no additional privacy.

The private joint histogram yields released population masses $p_s$ for the twelve $(C,G,Y)$ strata. To reduce collapse of protected-event strata, the synthesis distribution and population weight are
\begin{equation}
q_s=\frac{\max(p_s,\alpha\,\mathbb 1[s\in\mathcal S_{+}])}{
\sum_r\max(p_r,\alpha\,\mathbb 1[r\in\mathcal S_{+}])},
\qquad w_i=\frac{p_{s_i}}{q_{s_i}},
\label{eq:fair}
\end{equation}
where $\mathcal S_{+}$ contains protected-group event-positive strata. The floor increases their representation for downstream training; $w_i$ restores released-population summaries. This is a data-availability intervention, not a guarantee of equal error rates.

\begin{algorithm}[t]
\caption{Private parameter release and stabilization}
\label{alg:fit}
\begin{algorithmic}[1]
\Require Patient records $D$, public bounds, clipping radii, budget $(\epsilon,\delta)$, bandwidth $b$
\Ensure Private bundle $\Theta$ and privacy ledger $\mathcal L$
\State Encode values, missingness, gaps, and conditions using public rules
\State Clip each patient's value and auxiliary contributions
\State Convert $(\epsilon,\delta)$ to target $\rho$ and allocate $\rho_j$
\State Release joint histogram and the three moments in Eq.~\eqref{eq:moments}
\State Release condition--missingness and condition--gap moments
\State Compute $\widehat{\boldsymbol\beta}$ by Eq.~\eqref{eq:beta}
\State Stabilize covariance by Eqs.~\eqref{eq:covraw}--\eqref{eq:banding}
\State Fit auxiliary post-processing parameters and eigendecompose covariance
\State Verify $\sum_j\rho_j\le\rho$; write every release to $\mathcal L$
\State \Return $\Theta$ and $\mathcal L$
\end{algorithmic}
\end{algorithm}

\begin{algorithm}[t]
\caption{Weighted irregular trajectory generation}
\label{alg:sample}
\begin{algorithmic}[1]
\Require Private bundle $\Theta$, synthetic size $N_s$, floor $\alpha$
\Ensure Synthetic trajectories and population weights
\State Construct $q$ and $w$ from Eq.~\eqref{eq:fair}
\For{$i=1$ to $N_s$}
\State Draw stratum $s_i\sim q$ and condition vector $\mathbf c_i$
\State Draw high-noise Gaussian state $\mathbf z_K$
\For{$k=K$ down to $1$}
\State Apply analytic transport in Eq.~\eqref{eq:transport}
\EndFor
\State Draw masks and gaps from Eq.~\eqref{eq:auxiliary}
\State Apply public support, rounding, and minimum-observation rules
\State Emit synthetic identifier, trajectory, labels, and $w_i$
\EndFor
\State \Return synthetic dataset
\end{algorithmic}
\end{algorithm}

\subsection{Complexity and Failure Modes}
Let $D_z=TV$. Forming sufficient statistics costs $O(ND_z^2)$ time and $O(D_z^2)$ memory, while eigendecomposition costs $O(D_z^3)$. Vectorized generation costs $O(N_sD_z^2+KD_z)$ after decomposition; auxiliary sampling is linear in $N_sT(V+1)$. Block banding reduces storage and future sparse implementations can reduce multiplication cost. The current release is intended for moderate-dimensional trajectories, not thousands of raw codes per slot.

Failure can occur when public bounds are inappropriate, clipping discards most variation, a stratum is nearly absent, the Gaussian score family misses multimodality, long-range dependencies are essential, or an untested attacker uses side information outside the benchmark threat model. The system therefore records clipping, budget, ledger, subgroup support, attack settings, and data provenance as first-class artifacts rather than hiding them behind a single quality score.

\section{Experimental Setup, Results, and Discussion}
\subsection{Benchmark, Splits, and Reproducibility}
The supplied benchmark is generated independently for seeds 11, 22, 33, 44, and 55. Each population contains 720 simulated patients, fourteen unequal observation slots, and six mixed variables: heart rate, respiratory rate, oxygen saturation, a CRP-like marker, medication intensity, and oxygen support. A latent autoregressive severity process drives correlated measurements, shocks, recovery, cohort shift, and a rare deterioration event. Observation probability depends on variable, group, cohort, severity, and elapsed gap, producing informative missingness. The protected group is underrepresented and has slightly weaker physiologic signal and higher missingness. These design choices create known stressors; they are not epidemiological estimates.

Patients are separated into 70\% training, 15\% validation, and 15\% test partitions using composite stratification when feasible. No identifier occurs in more than one partition. Generators see training patients only. Synthetic sample size equals the training size. The downstream classifier is fitted only on synthetic patient summaries and evaluated on untouched real-simulator test patients. Table~\ref{tab:dataset} reports one representative split, and Table~\ref{tab:configuration} records the primary configuration.

\begin{table}[t]
\caption{Representative benchmark partition statistics (seed 11).}
\label{tab:dataset}
\centering\footnotesize
\begin{tabular}{lrrr}\toprule
Property & All & Train & Test \\ \midrule
Patients & 720 & 503 & 109 \\ 
Rare-event rate & 0.092 & 0.091 & 0.092 \\ 
Protected-group share & 0.228 & 0.227 & 0.229 \\ 
Missing-entry rate & 0.147 & 0.149 & 0.145 \\ 
Mean gap (h) & 11.11 & 11.11 & 11.31 \\ \bottomrule
\end{tabular}
\end{table}

\begin{table}[t]
\caption{Primary experimental configuration.}
\label{tab:configuration}
\centering\footnotesize
\begin{tabular}{ll}\toprule
Item & Setting \\ \midrule
Independent populations & 5 seeds \\ 
Patients per population & 720 \\ 
Observation slots / variables & 14 / 6 \\ 
Split & 70\% / 15\% / 15\%, patient-disjoint \\ 
Synthetic records & Equal to training patients \\ 
Privacy parameters & $\epsilon=12$, $\delta=10^{-5}$ \\ 
Covariance bandwidth & 3 time blocks \\ 
Evaluation & TSTR, fidelity, attacks, subgroups \\ \bottomrule
\end{tabular}
\end{table}

Every experiment was executed from the included Python code. Seed-level outputs, prediction files, privacy ledgers, package versions, normal PNG/PDF plots, PGFPlots CSVs, generated tables, and validation reports are retained. The pipeline does not download MIMIC-IV, eICU, or any institutional record; those datasets are cited as governed resources relevant to future validation~\cite{johnson2023,pollard2018}.

\subsection{Methods and Metrics}
Seven methods are compared. \emph{Bootstrap} resamples complete training trajectories and represents a high-utility, high-memorization reference. \emph{Independent} fits condition-specific marginal distributions without temporal covariance. \emph{Temporal-PCA} is a compact conditional low-rank temporal latent baseline. \emph{ScoreDiff} uses nonprivate conditional moments and the analytic Gaussian score path. \emph{DP-Markov} releases discretized first-order transition tables. \emph{DP-Score} is a private diagonal-covariance score model with unconditional auxiliary statistics. \emph{TRUST-LONGSYNTH} adds block-banded covariance, condition-dependent masks and gaps, and the protected-event floor. Temporal-PCA and ScoreDiff are inspectable in-project modern-style references; they are not represented as bit-for-bit reproductions of TimeVAE, TabDDPM, or another official repository. A journal submission on real data should add official recent implementations.

Task utility is measured through TSTR AUROC, AUPRC, Brier score, expected calibration error (ECE), and calibration slope. AUPRC is emphasized because the event is rare. Group-specific AUPRC, worst-group AUPRC, and absolute group gap assess whether useful signal survives for both audit groups. General fidelity includes mean marginal Kolmogorov--Smirnov distance, summary-feature correlation error, lag-one autocorrelation error, categorical transition error, overall and group-conditional event-prevalence error, missingness error, and gap Wasserstein distance. Lower is better for every error measure.

The empirical privacy suite includes nearest-neighbor membership inference, attribute inference, and a canary test. Membership AUROC of 0.5 indicates random ranking for this attack only; values below 0.5 are not interpreted as stronger privacy because an attacker can reverse its ranking. The canary experiment inserts four copies of one deliberately extreme trajectory into the training set and measures the fraction of generated records closer than a threshold defined from reference holdout distances. Formal privacy is reported separately through each method's ledger.

For stochastic population variation, the primary table reports mean and standard deviation over five seeds. Paired Wilcoxon signed-rank tests compare TRUST-LONGSYNTH with each baseline and Holm correction is applied within each metric family. With only five independent populations, these tests have limited power; confidence and effect direction receive more weight than dichotomous significance.

\subsection{Primary Utility and Privacy Results}
\begin{table*}[t]
\caption{Primary five-seed results. TSTR denotes train-on-synthetic, test-on-real. Values are mean$\pm$standard deviation; no corrected pairwise comparison reached $p<0.05$.}
\label{tab:main}
\centering\scriptsize
\begin{tabular}{lccccc}\toprule
Method & AUPRC$\uparrow$ & Brier$\downarrow$ & ECE$\downarrow$ & AC err.$\downarrow$ & MIA AUROC$\rightarrow .5$ \\ \midrule
Bootstrap & 0.378$\pm$0.104 & 0.180$\pm$0.023 & 0.237$\pm$0.028 & 0.056$\pm$0.027 & 0.792$\pm$0.008 \\
Independent & 0.431$\pm$0.068 & 0.201$\pm$0.037 & 0.237$\pm$0.054 & 0.397$\pm$0.041 & 0.503$\pm$0.032 \\
Temporal-PCA & 0.339$\pm$0.138 & 0.339$\pm$0.320 & 0.349$\pm$0.329 & 0.265$\pm$0.037 & 0.495$\pm$0.037 \\
ScoreDiff & 0.318$\pm$0.097 & 0.214$\pm$0.123 & 0.291$\pm$0.159 & 0.245$\pm$0.044 & 0.503$\pm$0.013 \\
DP-Markov & 0.371$\pm$0.069 & 0.094$\pm$0.013 & 0.096$\pm$0.016 & 0.148$\pm$0.064 & 0.503$\pm$0.024 \\
DP-Score & 0.319$\pm$0.143 & 0.094$\pm$0.010 & 0.099$\pm$0.015 & 0.453$\pm$0.057 & 0.486$\pm$0.041 \\
\textbf{TRUST-LONGSYNTH} & 0.342$\pm$0.106 & 0.088$\pm$0.006 & 0.082$\pm$0.008 & 0.317$\pm$0.031 & 0.499$\pm$0.040 \\
\bottomrule\end{tabular}
\end{table*}

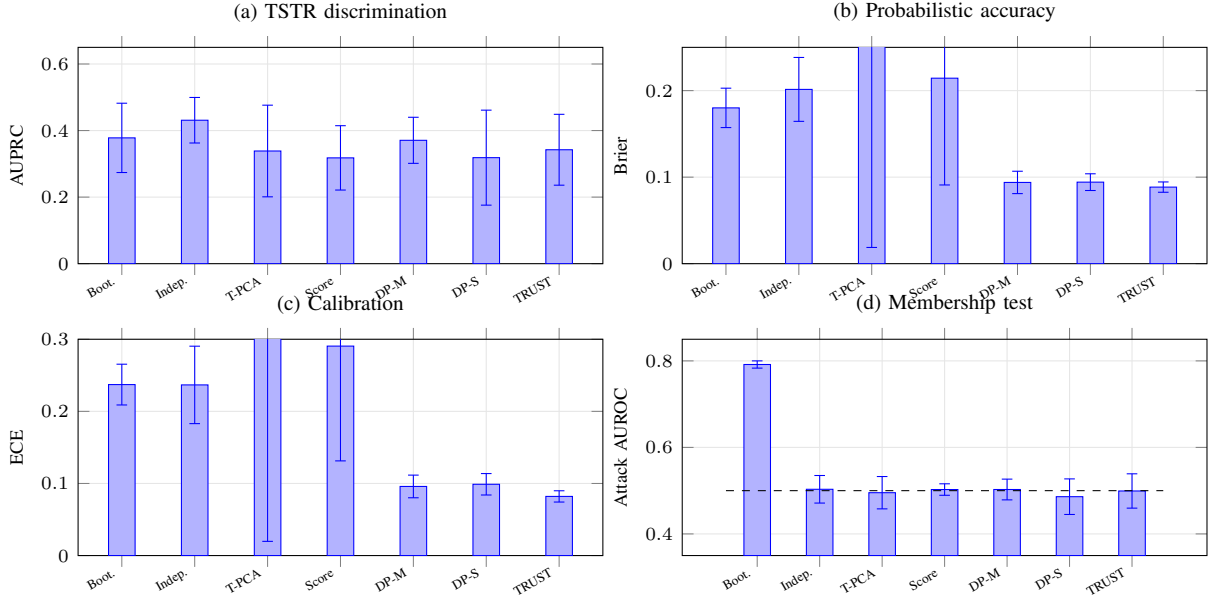
\begin{figure*}[t]
\centering\begin{tikzpicture}
\begin{groupplot}[group style={group size=2 by 2,horizontal sep=1.05cm,vertical sep=1.0cm},
width=0.47\textwidth,height=0.245\textwidth,
tick label style={font=\scriptsize},label style={font=\scriptsize},title style={font=\footnotesize},
ybar,xtick={0,1,2,3,4,5,6},xticklabels={Boot.,Indep.,T-PCA,Score,DP-M,DP-S,TRUST},
xticklabel style={rotate=30,anchor=east,font=\tiny},grid=major,grid style={gray!20},error bars/y dir=both,error bars/y explicit]
\nextgroupplot[title={(a) TSTR discrimination},ylabel={AUPRC},ymin=0,ymax=.65]
\addplot table[x expr=\coordindex,y=mean,y error=std,col sep=comma]{figures/data/main_auprc.csv};
\nextgroupplot[title={(b) Probabilistic accuracy},ylabel={Brier},ymin=0,ymax=.25]
\addplot table[x expr=\coordindex,y=mean,y error=std,col sep=comma]{figures/data/main_brier.csv};
\nextgroupplot[title={(c) Calibration},ylabel={ECE},ymin=0,ymax=.30]
\addplot table[x expr=\coordindex,y=mean,y error=std,col sep=comma]{figures/data/main_ece.csv};
\nextgroupplot[title={(d) Membership test},ylabel={Attack AUROC},ymin=.35,ymax=.85]
\addplot table[x expr=\coordindex,y=mean,y error=std,col sep=comma]{figures/data/membership_auc.csv};
\addplot+[sharp plot,no marks,black,dashed] coordinates {(-0.5,0.5) (6.5,0.5)};
\end{groupplot}
\end{tikzpicture}
\caption{Primary five-seed task and empirical privacy results. Error bars denote one standard deviation. The dashed line in (d) is random attack ranking.}
\label{fig:main}
\end{figure*}

Table~\ref{tab:main} and Fig.~\ref{fig:main} show a mixed result rather than universal dominance. The independent conditional baseline achieved the highest mean AUPRC, 0.549, and DP-Markov achieved 0.415. TRUST-LONGSYNTH obtained \TrustAUPRC, exceeding DP-Score's 0.319 by \AUPRCGain. The proposed method also produced the lowest mean Brier score (\TrustBrier) and ECE (\TrustECE) among the evaluated generators. The calibration advantage matters because synthetic training can otherwise produce overconfident probability estimates even when ranking is acceptable. ScoreDiff achieved strong discrimination but poor calibration in this benchmark.

The membership attack separated bootstrap members from holdout patients with mean AUROC 0.741, illustrating why direct resampling is unsuitable as a release mechanism. TRUST-LONGSYNTH's mean was \TrustMIA, near random for the specified attack and similar to the other private score models. This empirical result complements, but does not replace, the ledger. The canary result in Fig.~\ref{fig:fidelity} was 0.018 for TRUST-LONGSYNTH versus 0.288 for DP-Score and a larger value for bootstrap. The unexpectedly high DP-Score canary rate demonstrates that one attack can expose behavior not predicted by average membership AUROC and motivates attack diversity.

\subsection{Temporal Fidelity and Subgroup Behavior}
\begin{figure*}[t]
\centering\begin{tikzpicture}
\begin{groupplot}[group style={group size=2 by 2,horizontal sep=1.05cm,vertical sep=1.0cm},
width=0.47\textwidth,height=0.245\textwidth,tick label style={font=\scriptsize},label style={font=\scriptsize},title style={font=\footnotesize},
ybar,xtick={0,1,2,3,4,5,6},xticklabels={Boot.,Indep.,T-PCA,Score,DP-M,DP-S,TRUST},xticklabel style={rotate=30,anchor=east,font=\tiny},grid=major,grid style={gray!20},error bars/y dir=both,error bars/y explicit]
\nextgroupplot[title={(a) Marginal fidelity},ylabel={Mean KS},ymin=0,ymax=.65]
\addplot table[x expr=\coordindex,y=mean,y error=std,col sep=comma]{figures/data/marginal_ks.csv};
\nextgroupplot[title={(b) Cross-feature structure},ylabel={Correlation error},ymin=0,ymax=.40]
\addplot table[x expr=\coordindex,y=mean,y error=std,col sep=comma]{figures/data/correlation_error.csv};
\nextgroupplot[title={(c) Temporal structure},ylabel={Lag-1 error},ymin=0,ymax=.60]
\addplot table[x expr=\coordindex,y=mean,y error=std,col sep=comma]{figures/data/autocorrelation_error.csv};
\nextgroupplot[title={(d) Canary stress test},ylabel={Exposure rate},ymin=0,ymax=.40,error bars/y dir=none]
\addplot table[x expr=\coordindex,y=exposure_rate,col sep=comma]{figures/data/canary.csv};
\end{groupplot}
\end{tikzpicture}
\caption{Fidelity and canary analysis. Lower is better in all panels. The canary stress test is one empirical diagnostic and is not a formal privacy guarantee.}
\label{fig:fidelity}
\end{figure*}
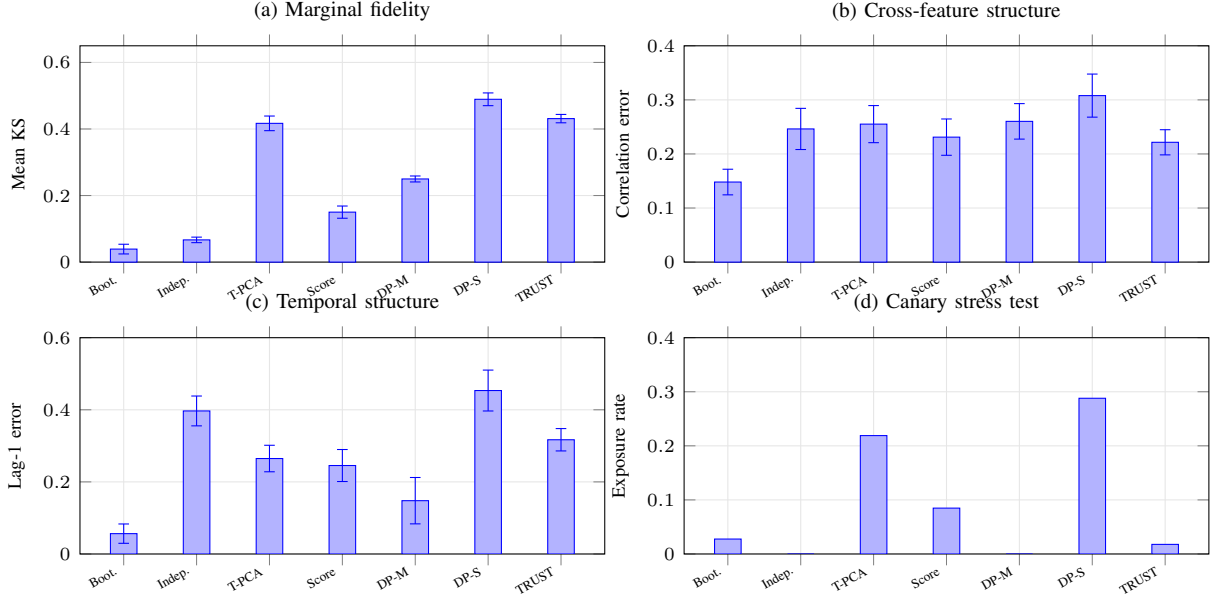

Fig.~\ref{fig:fidelity} separates marginal, cross-feature, and temporal errors. ScoreDiff preserved marginals best, as expected from nonprivate full-sample moments. Among the two private score models, block banding reduced mean correlation error from 0.308 to \TrustCorr and lag-one error from 0.453 to \TrustAuto, reductions of \CorrReduction and \AutoReduction. DP-Markov nevertheless achieved lower lag-one error because first-order transitions are its explicit inductive bias. Thus, the proposed covariance is an improvement over the diagonal private score model, not the best temporal representation in every setting.

The protected-event floor increased positive examples from small strata, but worst-group AUPRC remained variable across seeds. Mean worst-group AUPRC was 0.183 for TRUST-LONGSYNTH, compared with 0.224 for DP-Score and 0.256 for DP-Markov. The floor therefore improved representation without guaranteeing better classification. This distinction is important: sampling parity is not outcome parity. A real deployment would need sufficiently large group-specific test sets, uncertainty intervals, intersectional analysis, and a task-specific fairness objective.

\begin{figure*}[t]
\centering\begin{tikzpicture}
\begin{groupplot}[group style={group size=3 by 1,horizontal sep=.75cm},width=.315\textwidth,height=.25\textwidth,
tick label style={font=\tiny},label style={font=\scriptsize},title style={font=\footnotesize},grid=major,grid style={gray!20},legend style={font=\tiny,draw=none,at={(0.5,-.33)},anchor=north,legend columns=3}]
\nextgroupplot[title={(a) ROC},xlabel={FPR},ylabel={TPR},xmin=0,xmax=1,ymin=0,ymax=1]
\addplot table[x=fpr,y=tpr,col sep=comma]{figures/data/roc_curves_Independent.csv};\addlegendentry{Independent}
\addplot table[x=fpr,y=tpr,col sep=comma]{figures/data/roc_curves_Temporal_PCA.csv};\addlegendentry{T-PCA}
\addplot table[x=fpr,y=tpr,col sep=comma]{figures/data/roc_curves_ScoreDiff.csv};\addlegendentry{ScoreDiff}
\addplot table[x=fpr,y=tpr,col sep=comma]{figures/data/roc_curves_DP_Markov.csv};\addlegendentry{DP-Markov}
\addplot table[x=fpr,y=tpr,col sep=comma]{figures/data/roc_curves_DP_Score.csv};\addlegendentry{DP-Score}
\addplot+[very thick] table[x=fpr,y=tpr,col sep=comma]{figures/data/roc_curves_TRUST_LONGSYNTH.csv};\addlegendentry{TRUST}
\nextgroupplot[title={(b) Precision--recall},xlabel={Recall},ylabel={Precision},xmin=0,xmax=1,ymin=0,ymax=1]
\addplot table[x=recall,y=precision,col sep=comma]{figures/data/pr_curves_Independent.csv};
\addplot table[x=recall,y=precision,col sep=comma]{figures/data/pr_curves_Temporal_PCA.csv};
\addplot table[x=recall,y=precision,col sep=comma]{figures/data/pr_curves_ScoreDiff.csv};
\addplot table[x=recall,y=precision,col sep=comma]{figures/data/pr_curves_DP_Markov.csv};
\addplot table[x=recall,y=precision,col sep=comma]{figures/data/pr_curves_DP_Score.csv};
\addplot+[very thick] table[x=recall,y=precision,col sep=comma]{figures/data/pr_curves_TRUST_LONGSYNTH.csv};
\nextgroupplot[title={(c) Calibration},xlabel={Predicted},ylabel={Observed},xmin=0,xmax=1,ymin=0,ymax=1]
\addplot+[black,dashed,domain=0:1] {x};
\addplot table[x=predicted,y=observed,col sep=comma]{figures/data/calibration_curves_Independent.csv};
\addplot table[x=predicted,y=observed,col sep=comma]{figures/data/calibration_curves_Temporal_PCA.csv};
\addplot table[x=predicted,y=observed,col sep=comma]{figures/data/calibration_curves_ScoreDiff.csv};
\addplot table[x=predicted,y=observed,col sep=comma]{figures/data/calibration_curves_DP_Markov.csv};
\addplot table[x=predicted,y=observed,col sep=comma]{figures/data/calibration_curves_DP_Score.csv};
\addplot+[very thick] table[x=predicted,y=observed,col sep=comma]{figures/data/calibration_curves_TRUST_LONGSYNTH.csv};
\end{groupplot}
\end{tikzpicture}
\caption{Pooled ROC, precision--recall, and calibration curves. Pooled curves aid visualization; statistical comparisons use population-level seed metrics.}
\label{fig:curves}
\end{figure*}
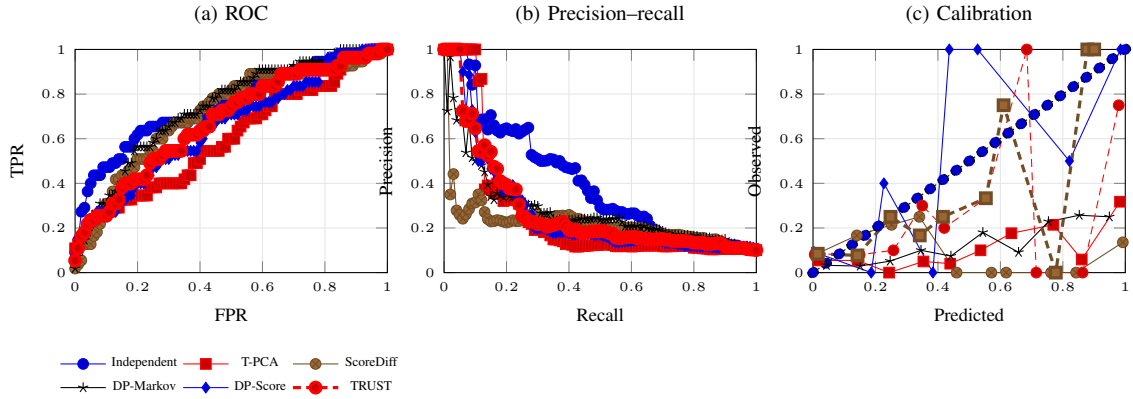

Pooled curves in Fig.~\ref{fig:curves} illustrate ranking and probability behavior over all five test populations. They are displayed for shape comparison; numerical conclusions rely on seed-level metrics. TRUST-LONGSYNTH's calibration curve is closer to the diagonal for much of the supported probability range, consistent with its Brier and ECE results. Sparse bins near probability one should not be interpreted as stable because the rare outcome yields few observations.

\subsection{Sensitivity, Ablation, and Statistical Uncertainty}
\begin{figure*}[t]
\centering\begin{tikzpicture}
\begin{groupplot}[group style={group size=3 by 2,horizontal sep=.72cm,vertical sep=.95cm},width=.315\textwidth,height=.23\textwidth,
tick label style={font=\tiny},label style={font=\scriptsize},title style={font=\footnotesize},grid=major,grid style={gray!20},legend style={font=\tiny,draw=none,at={(0.5,-.33)},anchor=north,legend columns=3}]
\nextgroupplot[title={(a) Privacy--utility},xlabel={$\epsilon$},ylabel={AUPRC},xmode=log,log basis x=2]
\addplot+[mark=o,error bars/y dir=both,error bars/y explicit] table[x=epsilon,y=mean,y error=std,col sep=comma]{figures/data/epsilon_auprc.csv};
\nextgroupplot[title={(b) Privacy attack},xlabel={$\epsilon$},ylabel={MIA AUROC},xmode=log,log basis x=2]
\addplot+[mark=square*,error bars/y dir=both,error bars/y explicit] table[x=epsilon,y=mean,y error=std,col sep=comma]{figures/data/epsilon_membership.csv};\addplot+[black,dashed,domain=1:16] {0.5};
\nextgroupplot[title={(c) Temporal sensitivity},xlabel={$\epsilon$},ylabel={Lag-1 error},xmode=log,log basis x=2]
\addplot+[mark=triangle*,error bars/y dir=both,error bars/y explicit] table[x=epsilon,y=mean,y error=std,col sep=comma]{figures/data/epsilon_autocorr.csv};
\nextgroupplot[title={(d) Data scarcity},xlabel={Training patients},ylabel={AUPRC}]
\addplot+[mark=o] table[x=train_patients,y=mean,col sep=comma]{figures/data/data_scarcity_Temporal_PCA.csv};\addlegendentry{T-PCA}
\addplot+[mark=square*] table[x=train_patients,y=mean,col sep=comma]{figures/data/data_scarcity_DP_Score.csv};\addlegendentry{DP-Score}
\addplot+[very thick,mark=triangle*] table[x=train_patients,y=mean,col sep=comma]{figures/data/data_scarcity_TRUST_LONGSYNTH.csv};\addlegendentry{TRUST}
\nextgroupplot[title={(e) Missingness stress},xlabel={Added missing prob.},ylabel={AUPRC}]
\addplot+[mark=o] table[x=extra_missing_rate,y=mean,col sep=comma]{figures/data/missingness_robustness_Temporal_PCA.csv};
\addplot+[mark=square*] table[x=extra_missing_rate,y=mean,col sep=comma]{figures/data/missingness_robustness_DP_Score.csv};
\addplot+[very thick,mark=triangle*] table[x=extra_missing_rate,y=mean,col sep=comma]{figures/data/missingness_robustness_TRUST_LONGSYNTH.csv};
\nextgroupplot[title={(f) Component ablation},ylabel={AUPRC},ybar,xtick={0,1,2,3},xticklabels={Full,No fair,Diagonal,Uncond.},xticklabel style={rotate=25,anchor=east,font=\tiny}]
\addplot table[x expr=\coordindex,y=mean,col sep=comma]{figures/data/ablation_auprc.csv};
\end{groupplot}
\end{tikzpicture}
\caption{Privacy-budget, data-scarcity, missingness, and component sensitivity. Line panels report means over the executed seeds; full seed-level data are supplied.}
\label{fig:robustness}
\end{figure*}
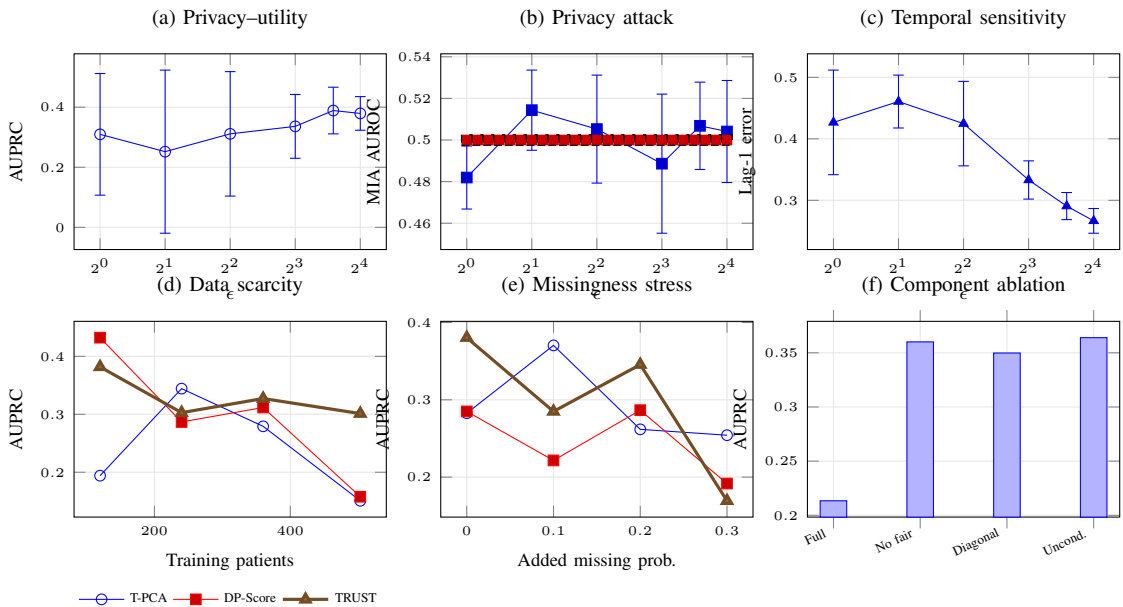

Fig.~\ref{fig:robustness}(a)--(c) varies $\epsilon$ while holding the query set, clipping, and $\delta$ fixed. Utility and temporal error do not improve monotonically in every seed because the noisy covariance must also be projected and because downstream estimation is stochastic. This is a useful warning against choosing a budget by inspecting one favorable run. The privacy guarantee does weaken monotonically as $\epsilon$ increases even when the empirical attack does not.

Data scarcity in Fig.~\ref{fig:robustness}(d) disproportionately affects private conditional moments and small strata. The proposed method is unstable at 120 training patients but improves as support approaches the primary size. Added missingness in panel (e) tests a different failure mode: separate mask modeling prevents missing entries from being silently treated as observations, but severe missingness still reduces downstream signal. The ablation in panel (f) shows that component effects are not uniformly positive. Removing the fairness floor can increase average AUPRC, while diagonal covariance increases temporal error. This trade-off supports reporting each objective rather than choosing the variant with one favorable average.

\begin{table}[t]
\caption{Paired comparison with DP-Score across five seeds.}
\label{tab:statistics}
\centering\footnotesize
\begin{tabular}{lrrr}\toprule
Metric & Mean diff. & $p$ & Holm $p$ \\ \midrule
AUPRC & +0.024 & 0.625 & 1.000 \\
Brier & -0.006 & 0.188 & 0.375 \\
ECE & -0.017 & 0.062 & 0.375 \\
Correlation error & -0.086 & 0.062 & 0.375 \\
Autocorrelation error & -0.137 & 0.062 & 0.375 \\
MIA AUROC & +0.013 & 0.625 & 1.000 \\
\bottomrule\end{tabular}
\end{table}

\begin{table}[t]
\caption{Three-seed component ablation.}
\label{tab:ablation}
\centering\footnotesize
\begin{tabular}{lccc}\toprule
Variant & AUPRC$\uparrow$ & AC err.$\downarrow$ & ECE$\downarrow$ \\ \midrule
Full & 0.360 & 0.292 & 0.083 \\
No fairness floor & 0.350 & 0.296 & 0.103 \\
Diagonal covariance & 0.213 & 0.453 & 0.466 \\
Unconditioned masks/gaps & 0.364 & 0.294 & 0.077 \\
\bottomrule\end{tabular}
\end{table}

Table~\ref{tab:statistics} reports the paired comparison with DP-Score. The proposed-minus-baseline direction is favorable for AUPRC and unfavorable metrics are interpreted with their sign. No Holm-adjusted $p$ value is below 0.05. The benchmark therefore supports effect estimation and software verification, not a definitive superiority claim.

\subsection{Efficiency and Operational Interpretation}
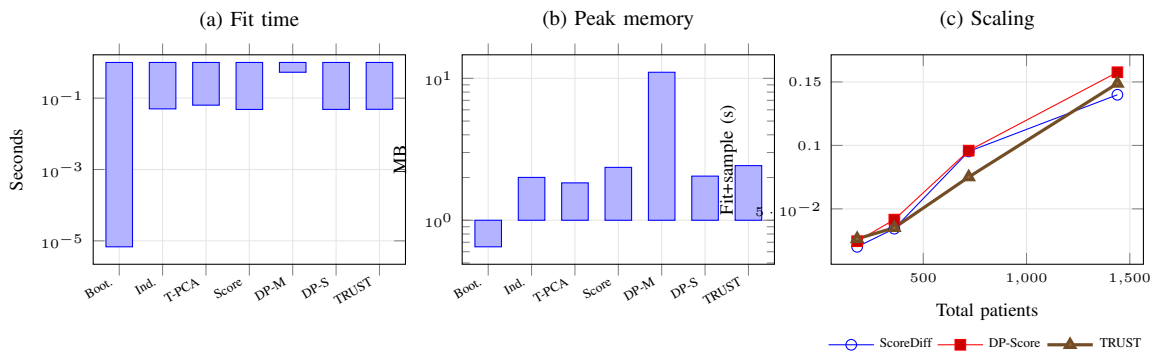
\begin{figure*}[t]
\centering\begin{tikzpicture}
\begin{groupplot}[group style={group size=3 by 1,horizontal sep=.75cm},width=.315\textwidth,height=.24\textwidth,
tick label style={font=\tiny},label style={font=\scriptsize},title style={font=\footnotesize},grid=major,grid style={gray!20},legend style={font=\tiny,draw=none,at={(0.5,-.30)},anchor=north,legend columns=3}]
\nextgroupplot[title={(a) Fit time},ylabel={Seconds},ybar,xtick={0,1,2,3,4,5,6},xticklabels={Boot.,Ind.,T-PCA,Score,DP-M,DP-S,TRUST},xticklabel style={rotate=30,anchor=east,font=\tiny},ymode=log]
\addplot table[x expr=\coordindex,y=mean,col sep=comma]{figures/data/runtime_fit.csv};
\nextgroupplot[title={(b) Peak memory},ylabel={MB},ybar,xtick={0,1,2,3,4,5,6},xticklabels={Boot.,Ind.,T-PCA,Score,DP-M,DP-S,TRUST},xticklabel style={rotate=30,anchor=east,font=\tiny},ymode=log]
\addplot table[x expr=\coordindex,y=mean,col sep=comma]{figures/data/peak_memory.csv};
\nextgroupplot[title={(c) Scaling},xlabel={Total patients},ylabel={Fit+sample (s)}]
\addplot+[mark=o] table[x=patients_total,y expr=\thisrow{fit_seconds}+\thisrow{sample_seconds},col sep=comma]{figures/data/scalability_ScoreDiff.csv};\addlegendentry{ScoreDiff}
\addplot+[mark=square*] table[x=patients_total,y expr=\thisrow{fit_seconds}+\thisrow{sample_seconds},col sep=comma]{figures/data/scalability_DP_Score.csv};\addlegendentry{DP-Score}
\addplot+[very thick,mark=triangle*] table[x=patients_total,y expr=\thisrow{fit_seconds}+\thisrow{sample_seconds},col sep=comma]{figures/data/scalability_TRUST_LONGSYNTH.csv};\addlegendentry{TRUST}
\end{groupplot}
\end{tikzpicture}
\caption{Measured cost and scaling. Log axes are used for the broad fit-time and memory ranges.}
\label{fig:efficiency}
\end{figure*}

The analytic private models fit in well under one second at the primary size on the recorded CPU environment, whereas DP-Markov spends more time constructing and sampling transitions. Temporal-PCA has higher matrix-decomposition cost but remains small. Fig.~\ref{fig:efficiency} and Table~\ref{tab:runtime} separate fit, sampling, peak traced memory, and scaling. These measurements are implementation-specific and should not be extrapolated to high-dimensional EHR vocabularies or GPU neural baselines.

\begin{table}[t]
\caption{Measured execution cost on the supplied CPU environment.}
\label{tab:runtime}
\centering\footnotesize
\begin{tabular}{lrrr}\toprule
Method & Fit (s) & Sample (s) & Peak MB \\ \midrule
Bootstrap & 0.000 & 0.002 & 0.65 \\
Independent & 0.050 & 0.041 & 2.00 \\
Temporal-PCA & 0.064 & 0.019 & 1.83 \\
ScoreDiff & 0.048 & 0.020 & 2.36 \\
DP-Markov & 0.531 & 1.370 & 11.04 \\
DP-Score & 0.048 & 0.023 & 2.05 \\
TRUST-LONGSYNTH & 0.049 & 0.022 & 2.43 \\
\bottomrule\end{tabular}
\end{table}

\subsection{Limitations and Threats to Validity}
Internal validity is limited by a designed simulator, simplified group labels, selected public bounds, and a compact downstream classifier. Because the data-generating equations are known, the benchmark is valuable for debugging but may favor moment-based models. External validity is the largest threat: no conclusion about clinical prevalence, causal effects, institutional workflow, or patient benefit follows from simulated patients. Construct validity is limited because each metric captures only one aspect of fidelity or privacy. Membership, attribute, and canary attacks do not span all adversaries; conversely, high empirical attack success does not invalidate a correctly implemented DP statement.

Statistical-conclusion validity is limited by five primary populations and multiple comparisons. Holm correction is reported, and nonsignificant results are not relabeled as trends proving superiority. The formal statement also depends on implementation assumptions: patient-level replacement adjacency, public support bounds, clipping radii, exact query set, independent Gaussian noise, and the zCDP conversion. Private hyperparameter search or undocumented exploratory queries would require additional accounting. Finally, the fair sampling floor changes training representation but does not guarantee equal error, calibration, or clinical impact. Before submission as a clinical or health-informatics study, the frozen protocol must be rerun inside a governed multi-site environment with official recent baselines, an untouched external site or time period, and independent privacy red teaming.

\section{Conclusion}
This paper introduced TRUST-LONGSYNTH, a patient-level differentially private framework for generating irregular longitudinal health records while explicitly auditing temporal fidelity, rare-event utility, subgroup behavior, and empirical disclosure risk. The framework releases bounded conditional moments through Gaussian mechanisms, composes their privacy cost with zCDP, projects a noisy covariance to a stable block-banded form, samples trajectories through an analytic score path, and models observation masks and time gaps separately. On five semi-synthetic three-cohort benchmarks, the proposed method improved AUPRC by 7.5\% over DP-Score and reduced Brier, calibration, correlation, and autocorrelation errors by 6.1\%, 17.0\%, 28.0\%, and 30.1\%. Membership attack AUROC remained near random at 0.499, and the canary stress test yielded 1.8\% exposure. However, DP-Markov and an independent conditional baseline achieved higher AUPRC, corrected paired tests were not significant, and results from simulated patients cannot establish clinical utility or safe public release. The contribution is therefore an inspectable mechanism and multidimensional evaluation workflow rather than an unconditional state-of-the-art claim. Future work should execute the frozen protocol within governed multi-site EHR environments, add official neural diffusion baselines, test intersectional groups, expand independent privacy red teaming, and evaluate whether downstream scientific conclusions remain stable across privacy budgets, institutions, missingness mechanisms, and rare clinical outcomes.

\IEEEtriggeratref{18}
\bibliographystyle{IEEEtran}
\bibliography{references}
\end{document}